\documentclass[intlimits,twoside,a4paper]{article}

\usepackage{amsmath}
\usepackage{amssymb}
\usepackage{graphicx}

\usepackage[T2A]{fontenc}
\usepackage[cp1251]{inputenc}

\usepackage{textcomp}

\usepackage{cmpj2}

\issue{2015}{18}{3}{33004}
\doinumber{10.5488/CMP.18.33004}
\title[Thermodynamics and kinetics of solids fragmentation]%
{Thermodynamics and kinetics of solids fragmentation at severe plastic deformation%
}
\author[A.V.~Khomenko, D.S.~Troshchenko, L.S.~Metlov]%
{A.V.~Khomenko\refaddr{label1}\thanks{E-mail: o.khomenko@mss.sumdu.edu.ua}\,, D.S.~Troshchenko\refaddr{label1}, L.S.~Metlov\refaddr{label2,label3}}
\addresses{
\addr{label1} Sumy State University, 2 Rimskii-Korsakov St., 40007 Sumy, Ukraine
\addr{label2} Donetsk Institute for Physics and Engineering named after O.O.~Galkin of the National Academy of Sciences of Ukraine, 46 Nauky ave., 03028 Kyiv, Ukraine
\addr{label3} Donetsk National University, 21 600-richchya St., 21021 Vinnytsia, Ukraine
}

\authorcopyright{A.V.~Khomenko, D.S.~Troshchenko, L.S.~Metlov, 2015}
\date{Received March 9, 2015, in final form June 22, 2015}

\begin{document}

\maketitle

\begin{abstract}
The approach of nonequilibrium evolution thermodynamics earlier offered is developed. It helps to describe the processes of defect formation within the adiabatic approximation. The basic equations system depends on the initial defects distribution (dislocations and grain boundaries). The phase diagram is determined with the domains of the realization of different limiting structure types. The interaction effect of several defect types on the formation of a limiting structure is investigated in terms of the internal energy. The conditions of the formation of two limiting structures are found. The kinetics of the steady-state values establishment of the defects density is investigated within the scope of the adiabatic approximation. The dislocations density change follows the evolution of the grain boundaries density in this approach. It is shown that grain sizes, in limiting structures, decrease with an increase of the elastic strains.
\pacs 05.70.-a, 
      61.72.Bb, 
      61.72.Mm, 
      62.20.F-, 
      64.60.-i, 
      68.35.Md  
\end{abstract}

\section{Introduction}\label{sec:level1}

Metals are subjected to different processing forms to achieve high mechanical properties (high \linebreak strength and plasticity). This purpose is the most cardinally reached by grinding the grain structure of metal as a result of its processing by the methods of severe (mega) plastic deformation (SPD) \cite{Segal, Valiev2000, Firstov2003, Beygelzimer, Goodarzy2014}. The microdisperse structure of grains with their linear sizes about 100~nm is formed by such processing. Since the processing is very complex and the real experiments are quite expensive, the development of theoretical methods for their description acquires a large significance.

Nowadays a few methods of the SPD processes description are known. The theory of fragmentation of the dislocation structures developed by G.A.~Malygin actually describes the formation of high-angular grain boundaries at large plastic deformations. This theory is based on the mechanism of deformation accompanied by dislocations self-organization \cite{Malygin2001, Malygin2002}. The flexural and twisting strains and stresses are formed during an equal channel angular extrusion process as a result of the generation of a large amount of geometrically required dislocations. However, having appeared their behavior does not differ from the behavior of chaotically arising dislocations. Actually, the author have extended the ideas developed earlier for the representation of the cellular structure formation \cite{Malygin1999}. These ideas are based on the description of the bigger defect generation (a grain boundary). It violates the hierarchical sequence of events because it would seem more logical to present the formation of grain boundaries as a result of the cellular structures self-organization rather than elementary dislocations. The newly formed structure, in the course of the $4$th and $5$th stages of material strengthening, should intersect a few elementary dislocation cells to save some integrity \cite{Malygin2001}. It is possible to assume that dislocations are more advantageous to abandon the volume of the dislocation cell or a grain and ``settle'' on their boundaries. Thereby they increase the level of their excessive energy (nonequilibrium) and only then form a new boundary like a process of its excessive energy relaxation.

The theory of material dispergation during the SPD process offered by V.I.~Kopylov and V.N.~Chu\-vil'deev \cite{Chuvildeev, Kopylov} is founded on the specific mechanism of deformation which is based upon the concept of interface slippage and anomalously high grain boundary diffusion. In addition, the representation of the free volume taken from the theory of amorphous materials is used in the present theory. At the same time, some supposition is put under a serious doubt by V.A.~Khonik experimental studies \cite{Khonik2004}. This is due to the fact that free volume can be used as a determining thermodynamic parameter for the description of amorphous alloys, at least in that classic interpretation accepted in the majority of the works. Indeed, the excess volume is the measure of an increase in the average distance that leads to the growth of the average potential energy of atoms. The distance between which is greater than the equilibrium one. However, an increase in the distance between certain atoms in a solid body, by virtue of the closure of force lines, inevitably entails diminishing of the distances between the other atoms (the local compression). Thus, the potential energy of the repulsion of atoms changes much sharper than the potential energy of weak Van der Waals attraction. Therefore, the potential energy at a lesser change of the volume due to the compression in this domain can substantially exceed the potential energy in the region of the excessive free volume. Consequently, it is very important to take into account in the theory not only the excessive free volume, but also ``the pressed volume'', which nobody has done so far. In addition, A.M.~Glezer presented in his lecture \cite{Glezer} that free volume has at least a bimodal structure in sizes. Amorphous materials having a more finely-structured free volume are more plastic, while the materials having a more coarse-structured free volume are more brittle. It is obvious that the relationships between free and pressed volume are different for them.

Besides, the role of actually diffusive mass transfer seems to be exaggerated. Really, from our point of view, under the action of excessive energy of the nonequilibrium boundary and internal stresses, a reconstruction of the material structure occurs, and grain boundaries actively participate herein. A boundary substantially facilitates such alteration providing additional degrees of freedom. Thus, mass transfer takes place, but it is not directly connected with the excess concentration of atoms.

Moreover, the theory \cite{Beygelzimer2005} originating from simple mechanical representations of the deformation processes is presently known. The system of kinetic equations is derived from the generalization of experimental results and laws. Within the framework of this theory, it is not possible to explain many important features of the SPD processes. It cannot describe the formation of the ``limiting'' structure of metals (minimum average grain sizes) in particular. Some attractor is heuristically introduced in the theory in order to describe the fragmentation of grains. It compels the system to aspire to the state with the required (fixed) grain sizes. Also the connection between the generation of several defect types has not been revealed yet, for example, such defects as the grain boundary and dislocation. Their interaction can provide the formation of stationary domains in the phase diagram.

At the same time, it is logical to suppose that the achievement of the limiting grain sizes as a result of dispergation during the SPD process is a general property of the steady-state material. It can be investigated in the most general form independently of the specific deformation mechanism. The general theory of the SPD processes was offered in the studies of the author \cite{Metlov_Bulletin2008,Metlov_PhysRevE2010,Metlov_PhysRevLett2011}, which is based on the nonequilibrium evolution thermodynamics. The formation of a limiting structure within the framework of this theory is related to a minimum or a maximum of the thermodynamic potential and is analogically described by the theory of phase transitions based on the general kinetic Landau-Khalatnikov-type equation \cite{Metlov_Bulletin2008,Metlov_PhysRevE2010,Metlov_PhysRevLett2011}. The main feature of this approach is that the defect structure is taken into account by an obvious introduction of the corresponding terms in the basic thermodynamic relationships which is similar to mesoscopic thermodynamics \cite{Rubi2003,2004condmat_Santamaria}, while the ordinary thermodynamics is an unstructured theory. The mesoscopic character of the model implies that large enough formations such as the grain, grain boundaries, accumulations of dislocations and so on can be the structural elements of the medium in such consideration \cite{Goodarzy2014}. Besides, special methods for the description of the behavior of a thin lubricant layer \cite{Khomenko2012,Lyashenko2011,Khomenko_TechPhys2005,ftt_2007,Khomenko_TechPhys2007} are being
developed at present. These methods are close to the concept of Landau theory of phase transitions.

The main purpose of this article is to describe the structural phase transition between two steady-states in terms of the internal energy (from the state with large grains to the small grain state). Further, the interaction effect of several defect types on the formation of a limiting structure should be investigated and the kinetics of the steady-state values establishment should be also described.

This article consists of four sections and the conclusions.

The basic theoretical methods of describing the evolution of the defect structure of  solids are given in  section~\ref{sec:level2}. Moreover, the basic relationship for the density of internal energy is written down. It combines the first law of thermodynamics and the law of energy transformation on the internal degrees of freedom. The two-defect two-mode model of the nonequilibrium evolution thermodynamics, in the expression of internal energy, is assumed to be the basis. The grain boundary and dislocation are chosen to be the main structural defects.

The phase diagram of fragmentation regimes by severe plastic deformation is studied in the section~\ref{sec:level3}. The evolution equations based on the expansion of internal energy have been derived at first. Then, using the adiabatic approach, the Landau-Khalatnikov equation is obtained. At last, the equation of steady-state values based on the necessary conditions of the extremum existence is found. Also, three graphs are built, which represent the dependence of the stationary values of the grain boundaries density on the second invariant of strain tensor, the phase diagram of the system, the dependence of the thermodynamic potential on the density of grain boundaries.

The kinetics of the density of the grain boundaries at the establishment of steady-state values is considered in  section~\ref{sec:level4}. This study includes three subsections. The relaxation dependencies are built here for the analysis of the Landau-Khalatnikov equation. The strain influence on the process of the equilibrium establishment in the system is examined in the second part. The check of adiabatic approximation used for the evolutionary system of equations of the density of defects is presented in the third subsection.

Short conclusions of our investigation are finally represented (section~\ref{sec:level5}).

\section{Basic equations}\label{sec:level2}

The conservation law of energy should be performed for both external interactions of the selected volume and the internal transformations of several energy types as a result of the flow of irreversible internal processes. Combining the first law of thermodynamics and the law of energy transformation in the internal degrees of freedom, it is possible to obtain thermodynamic ``identity'' for the density of internal energy $u$ as follows:
\begin{equation}
      \rd u=\sigma_{ij}\rd\varepsilon^\mathrm{e}_{ij}+T\rd s+\sum_{i=1}^N {\tilde{T}_l\delta \tilde{s}_l}+\sum_{i=1}^N \varphi_l\delta h_l\,,
      \label{eq1}                                              
\end{equation}
where $\sigma_{ij}$ and $\varepsilon_{ij}^\mathrm{e}$ are the stress tensor and elastic part of the strain tensor; $T$, $s$ and $\tilde{T}_l$, $\tilde{s}_l$ are a temperature and an entropy of the equilibrium and $l$-th nonequilibrium subsystems; $\varphi_l$ and $h_l$ are a conjugate pair of the thermodynamic variables that show the imperfection of a material (the energy and the defect density of $l$-type).

The nonequilibrium state is defined by a set of parameters. The first two $\varepsilon_{ij}^\mathrm{e}$ and $s$ describe the part of the system which has already come to an equilibrium distribution (the reversible processes), and the other two parameters $\tilde{s}_l$ and $h_l$ represent the nonequilibrium part (the irreversible processes) \cite{2013_arXiv3602}.

The relationship (\ref{eq1}) is written down in the general form. The specific model of the kinetics of structural defects will be determined, if we define the dependence of the internal energy or the effective internal energy on all independent variables of a problem \cite{2013_arXiv6791,Metlov}. Since the exact analytical solution is not known, let us consider a simplified model. We shall expand the effective internal energy into power series of its arguments. In this paper, the two-defect two-mode model is considered with the contribution of grain boundaries taking into account up to the fourth degree relatively to their density \cite{Metlov}.

The number of modes is determined by the number of stable stationary solutions or the maxima of internal energy. The number of levels is defined by the quantity of the considered types of defects. The grain boundary is the main structural defect during the SPD processes, but, at the same time, dislocations play an important role in the generation of power conditions for the formation of grain boundaries. Within the framework of this model during the SPD process, the main purpose is to describe the structural phase transition between two stable states, from a state with a large grain to the state with a small grain (100~nm) \cite{2013_arXiv6791,Metlov,Metlov_BulletinSerA2009}.

The internal energy is represented by the relationship \cite{Metlov_Bulletin2008,Metlov}:
\begin{equation}
      u\left( h_g, h_D \right) {=} u_0 {+}\sum_{m=g,D} \left( \varphi_{0m} h_m{-}\frac12 \varphi_{1m}h_m^2{+}\frac13 \varphi_{2m}h_m^3 {-}\frac14\varphi_{3m}h_m^4 \right) {+}\varphi_{gD} h_g h_D\,,  \label{eq2}                                             
\end{equation}
where $u_0$, $\varphi_{km}$, $\varphi_{gD}$ are some coefficients depending on the equilibrium variables $s$ and $\varepsilon_{ij}^\mathrm{e}$ as the control parameters:
\begin{eqnarray}
      &u_0=\frac12\lambda\left(\varepsilon_{ii}^\mathrm{e}\right)^2 +\mu\left(\varepsilon_{ij}^\mathrm{e}\right)^2, \label{eq3} \\
      &\varphi_{0m}=\varphi_{0m}^* +g_m\varepsilon_{ii}^\mathrm{e} +\left[\frac12\bar{\lambda}_m\left(\varepsilon_{ii}^\mathrm{e}\right)^2 +\bar{\mu}_m\left(\varepsilon_{ij}^\mathrm{e}\right)^2\right],
      \label{eq4} \\
      &\varphi_{1m}=\varphi_{1m}^* -2e_m\varepsilon_{ii}^\mathrm{e}\,, \label{eq5}   
\end{eqnarray}
where $\lambda$, $\mu$ are the elastic constants of the defect-free material; {$g_m$ is a positive constant characterizing the intensity of the defect production at $\varepsilon_{ii}^\mathrm{e}>0$ (comprehensive tension) or the defect annihilation or suppression of the defects generation at $\varepsilon_{ii}^\mathrm{e}<0$ (comprehensive compression); $\bar{\lambda}_m$, $\bar{\mu}_m$ are the elastic constants caused by the defects existence; $e_m$ is a positive constant responding similarly to $g_m$ for the defect production at $\varepsilon_{ii}^\mathrm{e}>0$ or for the defect annihilation at $\varepsilon_{ii}^\mathrm{e}<0$;} $\varepsilon_{ii}^\mathrm{e}$ and $\left(\varepsilon_{ij}^\mathrm{e}\right)^2 = \varepsilon_{ij}^\mathrm{e} \varepsilon_{ji}^\mathrm{e}$ are the first and the second invariants of the strain tensor. The repeated indexes mean summation. Since the compression of the deformed object is described, the negative values of the first invariant of the strain tensor $\varepsilon_{ii}^\mathrm{e}$ are used for further analysis.

The components of the strain $\varepsilon_{ij}^\mathrm{e}$ are the control parameters, which represent the external influence, and they can be regarded as constants. Index $D$, in equation~(\ref{eq2}), belongs to dislocations, and index $g$ belongs to grain boundaries.

{Note that the magnitude of the plastic strain, in the presented theory, does not appear in an explicit form.
However, it is present in a latent form. It is known that the accumulated strain is directly proportional to time at the constant deformation rate, and monotonously depends on time at variable rate of the deformation. Therefore, the accumulated strain is accepted to be used instead of time in the mechanics (dead time). In the presented theory, time is used in an explicit form, and the plastic strain appears in the latent form, namely, in the form of the defect density (see the derivation of a generalized Gibbs relation in \cite{Metlov_PhysRevE2014}).}

A polynomial of fourth degree with positive coefficients $\varphi_{km}$, in equation~(\ref{eq2}), can have two maxima (two modes). We shall consider only a simplified instance of the homogeneous distribution of dislocations. Therefore, the highest powers are neglected at the description of the dislocations evolution $\varphi_{2D}$ and $\varphi_{3D}$ \cite{Metlov}.

The following set of parameters is accepted for calculations:
\begin{align*}
    \varphi_{0g}^*&=5\cdot10^{-3}\:\textrm{J}\cdot \textrm{m}^{-2},& g_g&=9{.}1\:\textrm{J}\cdot \textrm{m}^{-2}, & \bar{\lambda}_g&=960\:\textrm{J}\cdot \textrm{m}^{-2},& \bar{\mu}_g &= 105\:\textrm{J}\cdot \textrm{m}^{-2},\\
    \varphi_{1g}^*&=3{.}3\:\textrm{J}\cdot \textrm{m}^{-1}, & e_g&=15{.}5\:\textrm{J}\cdot \textrm{m}^{-1}, & \varphi_{2g}&=6{.}5\:\textrm{J}, & \varphi_{3g}&=2{.}88\:\textrm{J}\cdot \textrm{m},\\
    \varphi_{0D}^*&=5\cdot10^{-4}\:\textrm{J}\cdot \textrm{m}^{-1},& g_D&=33{.}1\:\textrm{J}\cdot \textrm{m}^{-1}, &\bar{\lambda}_D&=96\:\textrm{J}\cdot \textrm{m}^{-1}, & \bar{\mu}_D &= 10{.}5\:\textrm{J}\cdot \textrm{m}^{-1},\\
    \varphi_{1D}^*&=0{.}6\:\textrm{J}\cdot \textrm{m}, & e_D&=1{.}55\:\textrm{J}\cdot \textrm{m}, & \varphi_{gD}&=0{.}03\:\textrm{J}. & &
\end{align*}

\section{Phase diagram}\label{sec:level3}

Let us write down the evolution equation \cite{Metlov_Bulletin2008, Metlov_PhysRevE2010, Metlov_PhysRevLett2011, Metlov}:
\begin{equation}   
      \tau_{h_l}\frac{\partial h_l}{\partial t}=\frac{\partial \bar{u}}{\partial h_l}\,,
      \label{eq6}
\end{equation}
where $\tau_{h_l}$ is the time of relaxation; $h_l$ is the density of $l$-type defects; $\bar{u}$ is the effective internal energy \cite{Metlov}. The system of evolution equations is defined in the explicit form \cite{Metlov}:
\begin{align}   
      \tau_{h_D}\frac{\partial h_D}{\partial t}&= \varphi_{0D}-\varphi_{1D}h_D + \varphi_{gD}h_g\,,
      \label{eq7} \\
      \tau_{h_g}\frac{\partial h_g}{\partial t}&= \varphi_{0g}-\varphi_{1g}h_g +\varphi_{2g}h_g^2 -\varphi_{3g}h_g^3 + \varphi_{gD}h_D\,.
      \label{eq8}
\end{align}

Let us use the adiabatic approximation $\tau_{h_g}\gg \tau_{h_D}$. The evolution of the density of dislocations follows the change of the density of grain boundaries under this condition. In this instance, we set $\tau_{h_D} \left({\partial h_D}/{\partial t}\right) \approx 0$ in equation~(\ref{eq7}) and express $h_D$ from this equation:
\begin{equation} \label{eq9}          
      h_D= \frac{\varphi_{gD}}{\varphi_{1D}}h_g+ \frac{\varphi_{0D}}{\varphi_{1D}}\,.
\end{equation}

Substituting the dependence of the density of dislocations (\ref{eq9}) into equation~(\ref{eq8}), the Landau-Khalat\-nikov equation is obtained:
\begin{equation} 
      \tau_{h_g}\frac{\partial h_g}{\partial t}=\frac{\partial V}{\partial h_g}\,, \label{eq10}
\end{equation}
where the derivative of the effective thermodynamic potential with respect to the density of grain boundaries  ${\partial V}/{\partial h_g} \equiv F\left(h_g\right)$ specifies the thermodynamic force $F$:
\begin{equation}   
      F\left( h_g\right) = \varphi_{0g} + \varphi_{gD}\frac{\varphi_{0D}}{\varphi_{1D}} - \left( \varphi_{1g} - \frac{\varphi_{gD}^2}{\varphi_{1D}}\right) h_g +\varphi_{2g}h_g^2 -\varphi_{3g}h_g^3\,,\label{eq11}
\end{equation}
that tends to bring the parameter $h_g$ to the attractor corresponding to the steady-state value. The system is described by the thermodynamic potential:
\begin{equation}    
     V\left( h_g \right) =\int\limits_0^{h_g} F\left( h_g' \right) \rd h_g'\,.\label{eq12}
\end{equation}
Using the substitution (\ref{eq9}), this relationship is identical to equation~(\ref{eq2}) for the given defect types.

The steady-state density of grain boundaries $h_g$ is fixed by the extremum condition of the potential (\ref{eq12}), since at ${\partial V}/{\partial h_g}=0$ according to equation~(\ref{eq10}) ${\partial h_g}/{\partial t}=0$. Besides, the minima of the potential correspond to the unstable states, but its maxima meet the stable states \cite{Metlov_Bulletin2008,Metlov_PhysRevE2010,Metlov_PhysRevLett2011,2013_arXiv6791}.

\begin{figure}[!t]
      \centerline{\includegraphics[width=0.47\textwidth]{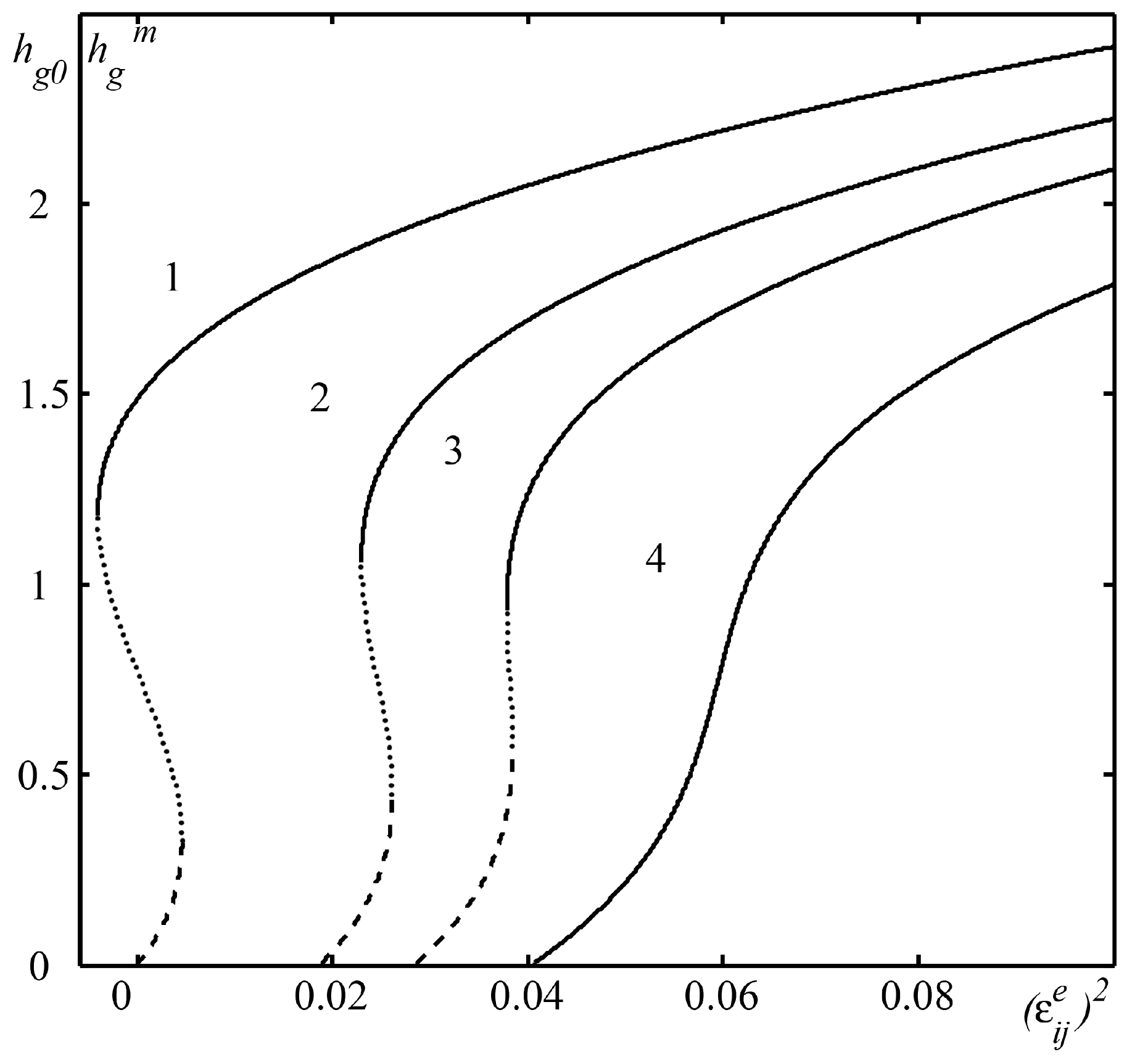}}
      \vspace{-2mm}
      \caption{\label{f1} Dependence of the stationary values of the density of grain boundaries $h_{g0}$, $h_g^m$ on the second
      invariant of the strain tensor $(\varepsilon_{ij}^\mathrm{e} )^2$. The curves 1--4 correspond to the
      values $\varepsilon_{ii}^\mathrm{e}$ $=0$, $-0{.}025$, $-0{.}042$, $-0{.}085$.} \label{fig1}
\end{figure}

The stationary condition leads to the expression:
\begin{equation}\label{eq13}    
 \varphi_{3g}h_g^3 - \varphi_{2g}h_g^2 + \left( \varphi_{1g} - \frac{\varphi_{gD}^2}{\varphi_{1D}}\right) h_g - \varphi_{0g} - \varphi_{gD} \frac{\varphi_{0D}}{\varphi_{1D}}=0\,.
\end{equation}
Hence, the positions of the potential extrema depend on the parameters
$\varphi_{0g}$, $\varphi_{1g}$, $\varphi_{2g}$, $\varphi_{3g}$, $\varphi_{0D}$, $\varphi_{1D}$ and do not depend on the reference level of the energy $u_0$. These extrema define the regimes of fragmentation during the SPD process. It is noteworthy that equation (\ref{eq13}) does not depend on the adiabatic approximation and is exact. This is related to the fact that solution is taken in the long-term asymptotic behavior when both of the stationary conditions are satisfied for both dislocations (\ref{eq9}) and grain boundaries.
Formal analysis of the solution of equation (\ref{eq13}) for different values of the control parameter $\varepsilon_{ii}^\mathrm{e}$
is presented in figure~\ref{fig1}. Note that we do not consider here the question of how to achieve this or that value of the control parameter (see \cite{2013_arXiv3602,2013_arXiv6791,Metlov,Metlov_BulletinSerA2009,Metlov_PhysRevE2014}). As seen, at small absolute values of the invariant $\varepsilon_{ii}^\mathrm{e}$, there exist three steady-states. Two of them correspond to the maxima of the potential $V\left( h_g\right)$ (solid and dashed lines) and one to the minimum of the potential (dotted line). The first maximum can be achieved at zero and non-zero values of the density of grain boundaries $h_{g0}$ depending on the value $(\varepsilon_{ij}^\mathrm{e} )^2$. It takes non-zero values only when the value of the strain $(\varepsilon_{ij}^\mathrm{e})^2$ is larger than a certain critical value. This is due to the fact that fragmentation process during the SPD is capable of occuring when the elastic strain $\varepsilon_{ij}^\mathrm{e}$ and the related to its stresses $\sigma_{ij}$ exceed the yield stress. The steady-states realized in the SPD process can be reached only at the fulfillment of this condition. If it fails, the system is also capable of approaching the stationary states but with another and lower rate\footnote{Let us note that with an increase of hydrostatic pressure, the limit of plastic flow also increases. The model correctly reflects the main regularities observed in real materials in this sense (see figure~13 in the article \cite{Demkowicz_PhysRevB2005}).}.

According to the curves 1--3, the smaller of the steady-states $h_{g0}$ corresponds to the grain with a large size (dashed parts of curves), the bigger of the steady-states (solid sections of curves) meets the fine grain. They are separated by unstable state (dotted line) for the values of the density of grain boundaries corresponding to the minimum of the potential. It is noteworthy that zero maximum meets the coarse-grained polycrystal or a single crystal in the limit. In the instance of a single crystal, the zero maximum of the potential is realized at first and only when it becomes non-zero, the process of fragmentation starts proceeding.

If the sample, before the SPD process, has already had the fine-grained structure (the state similar to the non-zero maximum of $V ( h_g)$ is realized), then according to the curve $1$, in figure~\ref{fig1}, the evolution of the material structure to the final steady-state is possible even for small values of $(\varepsilon_{ij}^\mathrm{e} )^2$.

According to the curve~1, with an increase of $(\varepsilon_{ij}^\mathrm{e} )^2$ under the value, when zero and non-zero maxima of the potential coexist, the process of fragmentation cannot be realized, because these maxima are separated by a potential barrier (dotted line). Then, zero maximum becomes non-zero (dashed line) and a continuous process of fragmentation occurs. At further increase of strain, the first maximum disappears coupled with the minimum and the system by the first-order phase transition rapidly passes into the state described by the second maximum of the potential (solid line). At the same time, an abrupt decrease of grain sizes takes place. It is known that at the first-order phase transition, the system can be in two metastable phases due to the simultaneous presence of two steady extrema of the thermodynamic potential \cite{Landau}.

If we continue to increase $\varepsilon_{ii}^\mathrm{e}$ for absolute magnitude (curve~4), the continuous transition, in the absence of the potential barrier, is realized from a single crystal to a fragmented sample. Besides, the formation of only one limiting structure is possible. In general, comparing the cures 1--4, we note an important fact that the application of hydrostatic pressure $\varepsilon_{ii}^\mathrm{e}$ restrains the generation of defects. It is necessary to apply a large shear strain $\varepsilon_{ij}^\mathrm{e}$ to generate them. The critical value of the second invariant of the strain tensor is obtained from equation~(\ref{eq13}) for $h_{g0} = 0$:
\begin{eqnarray}                 
     \left(\varepsilon_{ij}^\mathrm{e} \right)_c^2 &= &-\frac{1}{\left( \varphi_{1D} \bar{\mu}_g + \varphi_{gD} \bar{\mu}_D\right)} \bigg[ \left( \frac12\bar{\lambda}_g \varphi_{1D}+ \frac12\bar{\lambda}_D \varphi_{gD}\right) \left(\varepsilon_{ii}^\mathrm{e} \right)^2  \nonumber \\
     &&+ \left( g_g\varphi_{1D}+g_D\varphi_{gD}\right) \varepsilon_{ii}^\mathrm{e} +\left(  \varphi_{0g}^* \varphi_{1D}+\varphi_{0D}^* \varphi_{gD}\right)\bigg].
 \label{eq14}\end{eqnarray}
 The relation~(\ref{eq14}), in the coordinates $(\varepsilon_{ij}^\mathrm{e} )^2 - \varepsilon_{ii}^\mathrm{e}$, represents the second-order curve below which there exists a steady-state solution of equation~(\ref{eq13}) corresponding to the maximum of $V\left( h_g\right)$ at the point $h_{g0} = 0$. The curves, in figure~\ref{fig1}, start from the point (\ref{eq14}) on the abscissa axis. Therefore, the expression (\ref{eq14}) represents the value of the second invariant of the strain tensor at which the process of fragmentation begins. Since equation~(\ref{eq14}) contains the value $\varepsilon_{ii}^\mathrm{e}$, all the curves start from different points.

\begin{figure}[!t]
      \centerline{\includegraphics[width=0.48\textwidth]{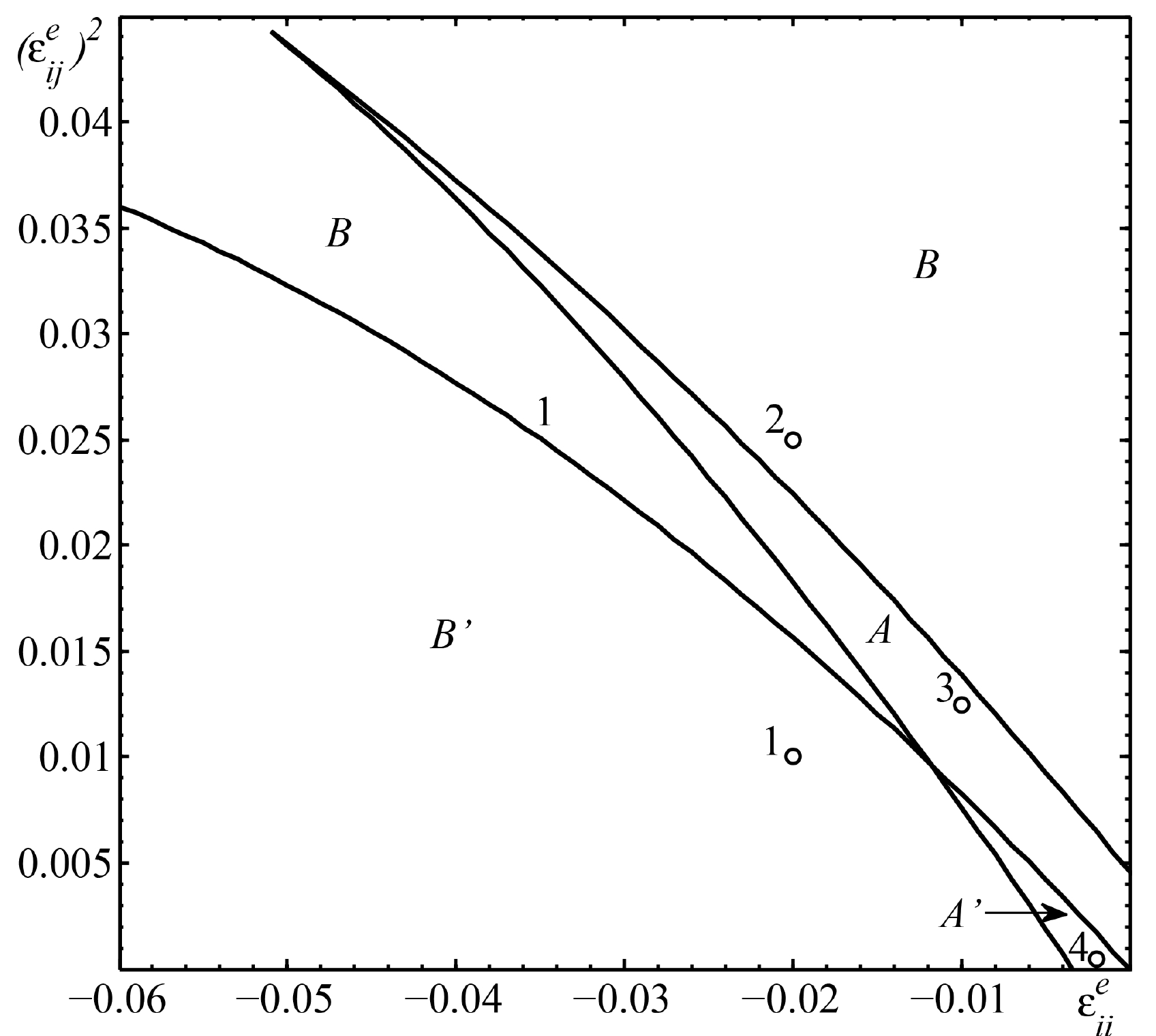}}
      \caption{\label{fig2} The phase diagram of the system with the realization of domains of two $\left( A,\:A'\right)$ and one $\left( B,\:B'\right)$ limiting structures.}
\end{figure}
\begin{figure}[!b]
      \centerline{\includegraphics[width=0.48\textwidth]{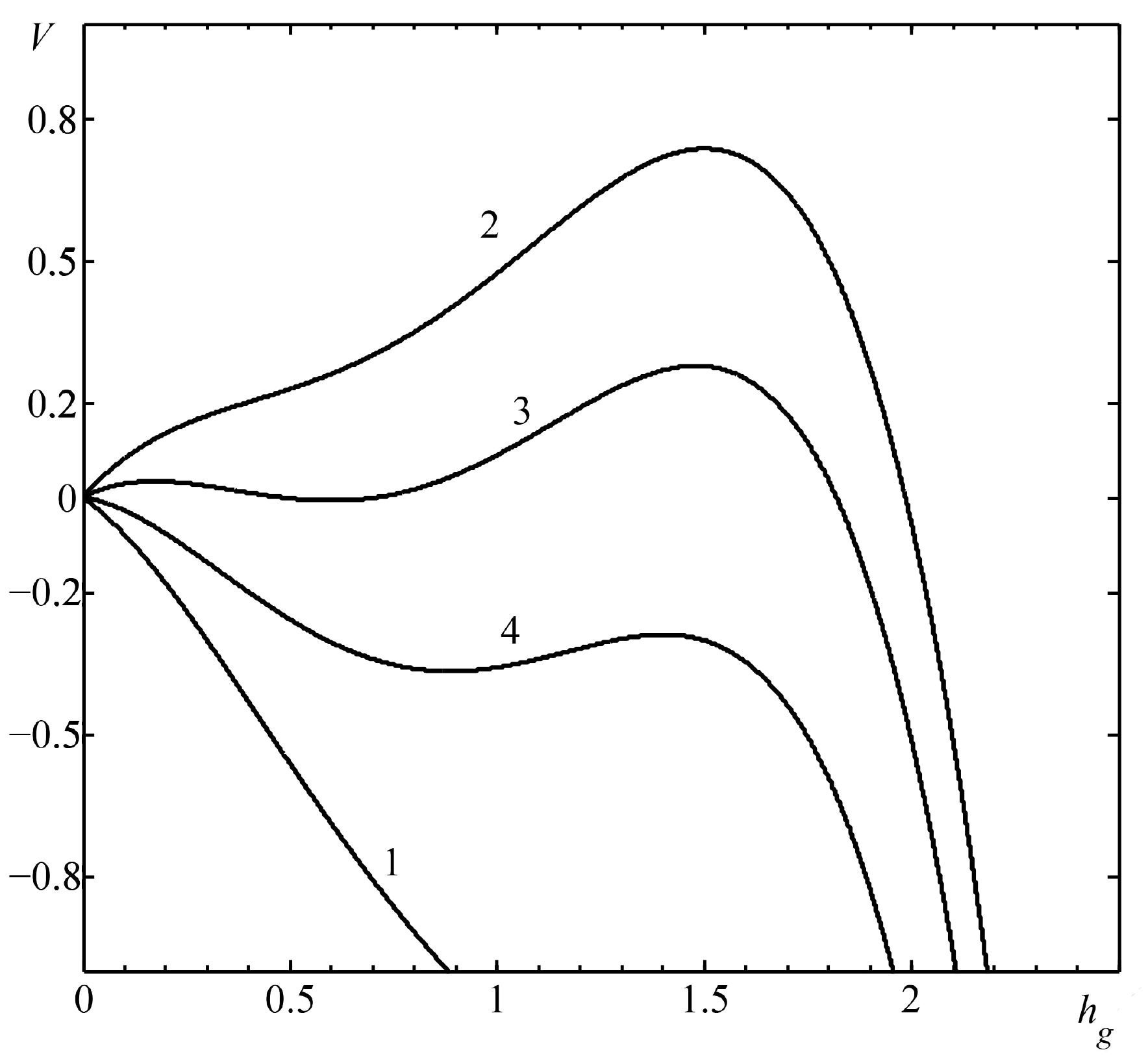}}
      \caption{\label{fig3} The dependence of thermodynamic potential $V\left( h_g\right)$ (\ref{eq12}) on the
grain boundaries density $h_g$. The curves 1--4 correspond to the values of the second invariant
      $(\varepsilon_{ij}^\mathrm{e} )^2 = 0{.}01,\:0{.}025,\:0{.}0125,\:0{.}0005$ and, accordingly, the first
      invariant $\varepsilon_{ii}^\mathrm{e} = -0{.}02,\:-0{.}02,\:-0{.}01,\:-0{.}002$ (points 1--4 in figure~\ref{fig2}).}
\end{figure}

The phase diagram is depicted in figure~\ref{fig2}. The lines correspond to the loss of the system stability. Curve 1 is defined by expression (\ref{eq14}), below which the zero steady-state solution is possible. There is not channel of energy dissipation for the value $h_{g0} = 0$, which is related with the formation of the defect structure, and the system is a single crystal or a structure close to it. The points 1--4, in the phase diagram, correspond to the potential curves, in figure~\ref{fig3}, which possess the maxima. Their positions are defined by the problem parameters.

The domain $A$ corresponds to the realization of two non-zero maxima of the potential $V\left( h_g\right)$ (curve~3 in figure~\ref{fig3}). Here, two limiting structures are observed with large (the first maximum of the potential) and small (the second maximum of the potential) grain sizes.

The region $A'$ of the diagram is similar to the domain $A$, but with the main difference that the first maximum of the potential is zero (the curve~4 in figure~\ref{fig3}). Since the first limiting or the stationary structure is formed for the value $h_{g0} = 0$, it is a single crystal. As a result of the SPD process, fragmentation of the material in this region cannot be realized. It is worth noting that transitions between the maxima of the potential are directly possible during the SPD processing. In the regions $A$ and $A'$, two limiting structures corresponding to different grain sizes are formed due to these processes. When the SPD process finishes, it should be supposed that the sample has been  formed and the further transitions are not realized.

According to curve~2, in figure~\ref{fig3}, one limiting structure is generated in the domain of a large strain $B$. It is seen that with an increase of elastic strain $(\varepsilon_{ij}^\mathrm{e} )^2$ the grain size decreases and the sample, in the limit $(\varepsilon_{ij}^\mathrm{e} )^2 \rightarrow \infty$, represents an amorphous structure. The only zero maximum of $V\left( h_g\right)$ (curve~1 in figure~\ref{fig3}) is realized in the domain of small strains $B'$. Here, the system is a single crystal.

{Note that the first maximum of thermodynamic potential $V\left( h_g\right)$ (\ref{eq12}) directly depends on the applied stresses, while the second one is not sensitive to the value of the first $\varepsilon_{ii}^\mathrm{e}$ and the  second $(\varepsilon_{ij}^\mathrm{e} )^2$ invariants of the elastic strain tensor. This is connected with the accuracy of taking into account the elastic strains, in the power representation of internal energy (\ref{eq2})--(\ref{eq5}). The considered strains are inserted only up to the 2-nd power of the corresponding density of defects $h_{m}$. They considerably effect the formation of the first maximum. Such an account is realized by virtue of preservation of the approximately identical order in all powers in relation to the internal energy (\ref{eq2}). The total order of the first two contributions of the density of defects $h_{m}$ and elastic strains $\varepsilon_{ii}^\mathrm{e}$, $(\varepsilon_{ij}^\mathrm{e})^2$ is approximately equal to the order of the last two ones, which do not take into account the strains. However, if the experiment shows the high sensitivity of the second maximum to the elastic strain, such a behavior can be considered by larger powers of approximation over $\varepsilon_{ii}^\mathrm{e}$ and $(\varepsilon_{ij}^\mathrm{e})^2$ in the coefficients $\varphi_{2m}$ and $\varphi_{3m}$ presented in relation to the internal energy (\ref{eq2}).}

\section{The kinetics of the steady-state values establishment of the density of grain boundaries}\label{sec:level4}

\subsection{Time dependencies}\label{subsec:level4_1}

Let us study the kinetics of the considered system and distinguish the simplified version of the system evolution within the scope of the adiabatic approximation $\tau_{h_g}\gg \tau_{h_D}$ \cite{Khomenko2012,Khomenko_MPAT2008,Khomenko_jfd_2007}. In so doing, the research is reduced to the analysis of the Landau-Khalatnikov kinetic equation (\ref{eq10}). Its explicit form is expressed as:
\begin{equation} 
     \dot{h}_g= \varphi_{0g} + \varphi_{gD} \frac{\varphi_{0D}}{\varphi_{1D}} - \left( \varphi_{1g} - \frac{\varphi_{gD}^2}{\varphi_{1D}}\right) h_g +\varphi_{2g}h_g^2 -\varphi_{3g}h_g^3\,.\label{eq15}
\end{equation}

\begin{figure}[!t]
\centerline{\includegraphics[width=0.65\textwidth]{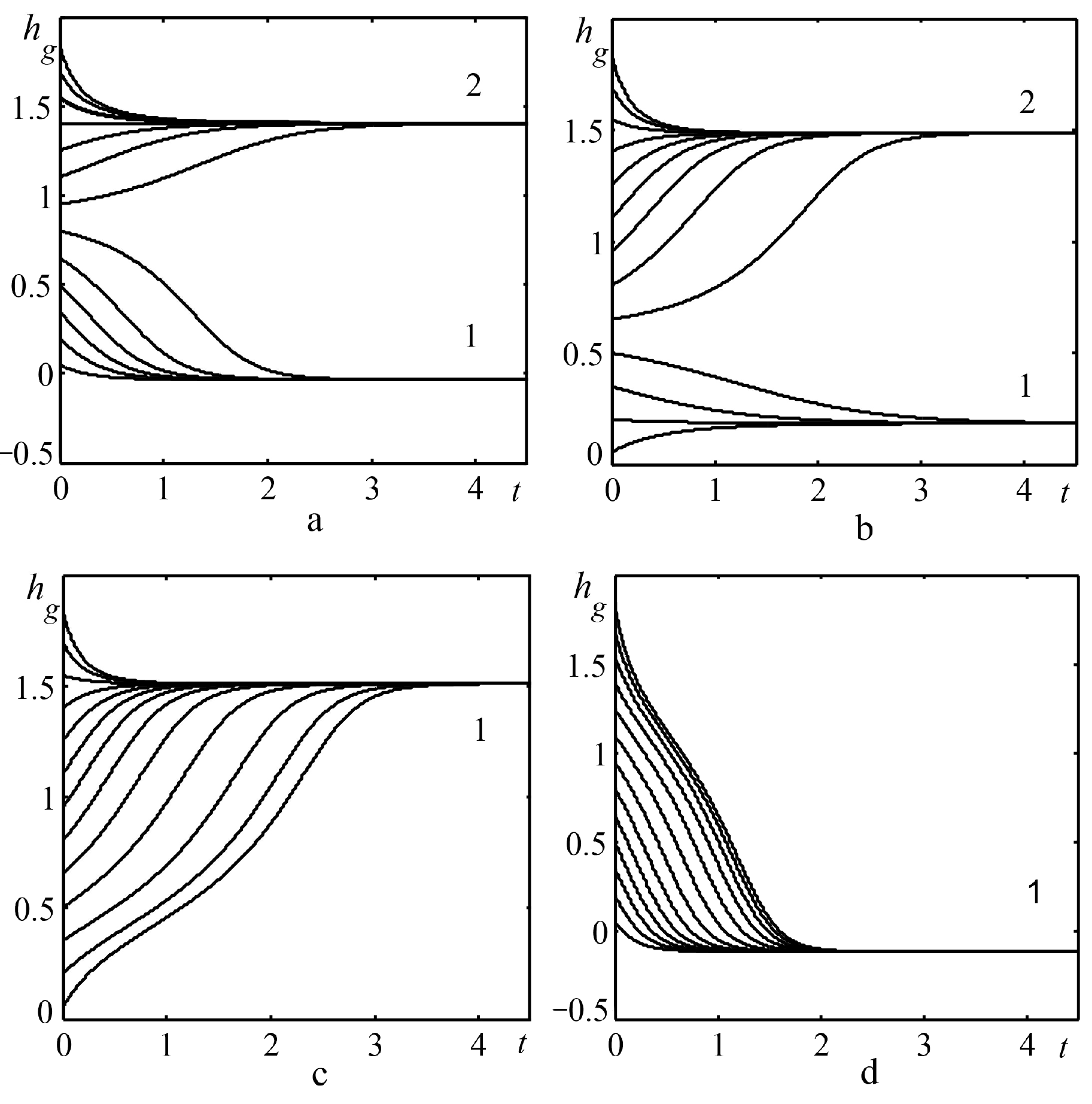}}
\caption{\label{fig4} Relaxation dependencies $h_g(t)[1/\textrm{m}]$ obtained by the solution of differential
equation (\ref{eq15}) for parameters of figure~\ref{fig2}. The figures (a)--(d) correspond to the points
4--1 of the phase diagram. The time $t[\textrm{s}]$ is measured in the unit of
$\tau_{h_g}[\textrm{J}\cdot \textrm{m}^{-1}\cdot \textrm{s}]$.}
\end{figure}

The relaxation time dependencies $h_g(t)$ are obtained by means of an approximate numerical solution of the differential equation (\ref{eq15}) (figure~\ref{fig4}) with corresponding parameters of figure~\ref{fig2}. Figures~\ref{fig4}~(a)--(d) are built for different domains of the phase diagram. The points 1--4 of the phase diagram (figure~\ref{fig2}) are chosen as the main parameters. It should be noted that the obtained time dependencies for equation (\ref{eq15}) within the scope of the adiabatic approach $\tau_{h_g}\gg \tau_{h_D}$ completely coincide with the solutions of system of equations (\ref{eq7}) and (\ref{eq8}). This suggests that both solutions are stable and  the usage of equation (\ref{eq15}), in the corresponding condition, is valid.

The dependencies describing the process of material fragmentation in the region $A'$ of the phase diagram are presented in figure~\ref{fig4}~(a). All curves are built for different initial conditions $h_g(t=0)$. The corresponding potential is shown by the curve~4, in figure~\ref{fig3}. The first stationary state (for a small initial value of $h_g(t=0)$) is in the negative region of $h_g$. This is due to the fact that the first maximum of the potential is actually realized for $h_g<0$. Since the negative values of $h_g$ have no physical meaning, so we suppose that the system is in the regime of $h_g=0$. It is considered that the density of grain boundaries stops decreasing and the system transforms into the corresponding state after zero value is achieved.

According to figure~\ref{fig4}~(a), both a single crystal (the system goes into the steady-state value that is described by the curve~1) and the fine-grained limiting structure (curve~2) can be realized depending on the initial conditions.

Time dependencies corresponding to the domain $A$ of the phase diagram (point~3 in figure~\ref{fig2}) are shown in figure~\ref{fig4}~(b). The main difference from the previous instance is that the first stationary state here is in the positive region of the density of grain boundaries $h_g$. In other words, two non-zero maxima of the potential $V\left( h_g\right)$ are realized (the curve~3 in figure~\ref{fig3}). Accordingly, two limiting structures with coarse (curve~1) and finer (curve~2) grains are formed.

The domain of large strains $B$ of diagram (figure~\ref{fig2}) is represented in figure~\ref{fig4}~(c). There is one steady state (curve~1). It corresponds to the one maximum of the potential $V\left( h_g\right)$ (curve~2 in figure~\ref{fig3}). The system relaxes to this value under any initial conditions and the limiting structure with the corresponding grain sizes is generated.

Figure~\ref{fig4}~(d) conforms to the domain of small strains $B'$ (point~1 in figure~\ref{fig2}). As the maximum of the potential $V\left( h_g\right)$ is observed at $h_g<0$, the steady-state is formed in the negative range of $h_g$. The value of $h_{g0}$ should be equal to zero, as well as for figure~\ref{fig4}~(a). Thus, as a result of the fast system relaxation to the steady-state, coarse single-crystal grains are formed. The corresponding potential is shown by curve 1 in figure~\ref{fig3}.

It should be noted that the first two terms, in equation (\ref{eq15}), describe some constant source of defects. This source leads to an increase in the density of grain boundaries $h_g$.
Large values of the first two terms move the steady-states of the system to the region of higher values of imperfection and promote the formation of finer grains. In particular, the existence of this source is connected with the material imperfection, namely, grain boundaries and dislocations, which during the deformation process interact with each other. Meanwhile, this does not rule out the presence of the defects of the deeper structural levels, such as impurity, inclusions of the other phase, a dislocation substructure, etc. For example, in case of alloys, the dependence of the limiting grain sizes on the percentage of the alloying elements is defined by this parameter. Thus, data are provided for the alloys on the basis of aluminum in the A.A.~Mazilkin et al. experimental work \cite{Mazilkin_PhysSS2007}. It follows from the data that the average size of the grain, in the alloy of Al--Mg, amounts to $150$ and $90$ nanometers for $5$ and $10\%$ of Mg; for alloys Al~--- $5\%$ Zn~--- $2\%$ Mg and Al~--- $10\%$ Zn~--- $4\%$ Mg the size of a  grain is respectively equal to 150 and 120 nanometers (before deformation 500 microns). This confirms the regularity stated above.

\subsection{Dependence of the \texorpdfstring{$\dot{h}_g$}{h'g} on \texorpdfstring{$h_g$}{hg}}\label{subsec:level4_2}

\begin{figure}[!h]
\centerline{\includegraphics[width=0.5\textwidth]{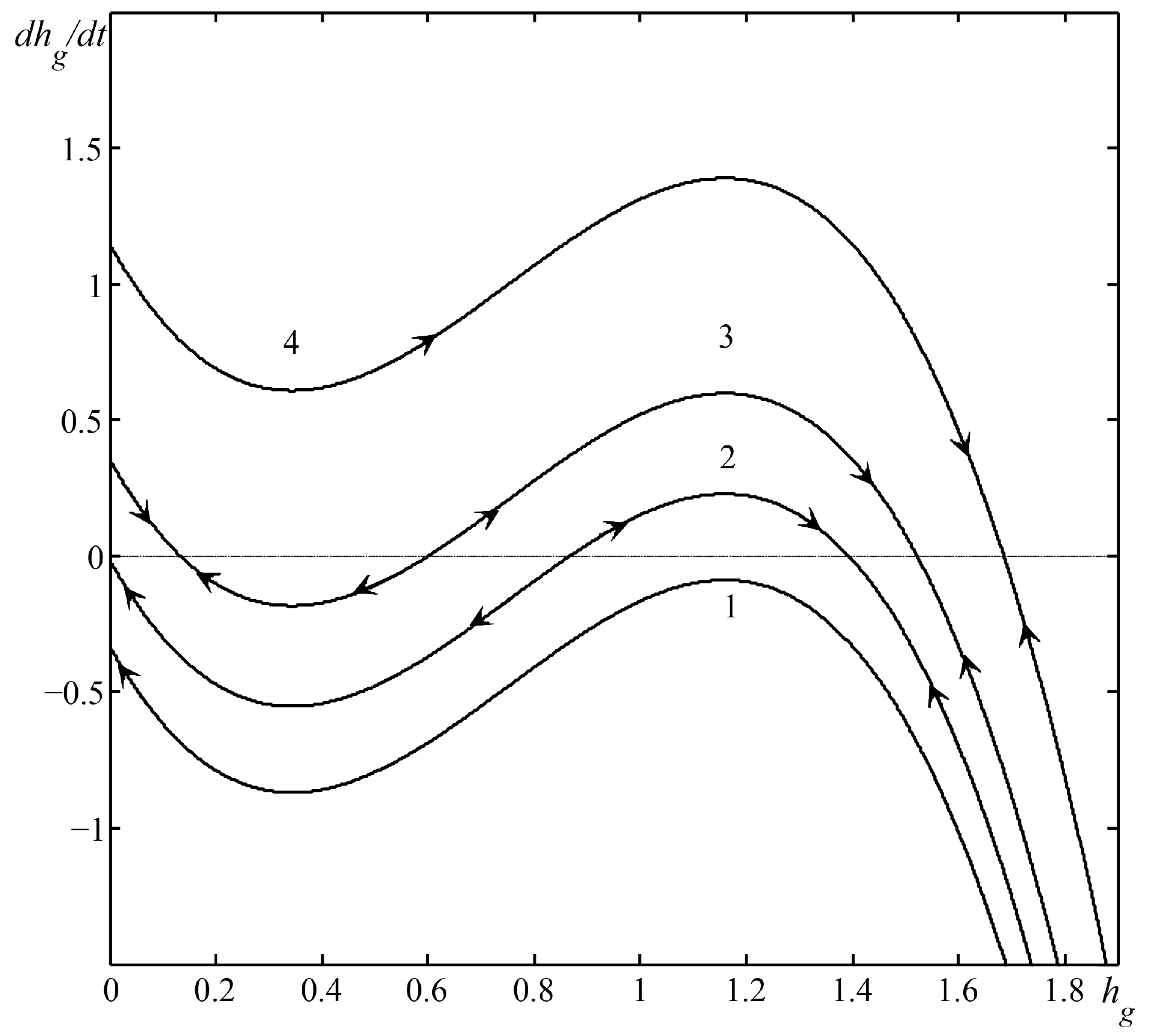}}
\caption{\label{fig5} The change rate of the density of grain boundaries $\dot{h}_g$ (\ref{eq15}) as a function of $h_g$ for the parameters of figure~\ref{fig1}. The curves $1$--$4$ correspond to the values of the first and the second invariants $\varepsilon_{ii}^\mathrm{e} =-0{.}005$ and $(\varepsilon_{ij}^\mathrm{e} )^2 =0{.}001,\:0{.}004,\:0{.}0075,\:0{.}015$.}
\end{figure}

Let us investigate the strain effect on the establishment process of the system equilibrium taking into account definitions (\ref{eq3})--(\ref{eq5}) \cite{Khomenko2004140,Khomenko_UJP2009}. The analyzed function is on the right-hand  side of equation (\ref{eq15}) [or $F\left( h_g\right)$ (\ref{eq11})] and is presented in figure~\ref{fig5}. The values of the first and the second invariants of the strain tensor for the construction of curves 1--4 are located in all the above mentioned domains of the phase diagram.

 Curve~1 is constructed at the smallest value of strain. It does not have the intersection point with the abscissa axis. As is seen in the figure, the change velocity of the density of grain boundaries is negative for the value $h_g=0$. This is due to the fact that the potential $V\left( h_g\right)$ has a maximum in the region of $h_g<0$. Since the change rate of grain boundaries is negative, it is supposed that the process of grain shattering is slowed down. However, the negative values of $h_g$ have no physical sense, because the channel of the energy dissipation (which is connected with the formation of defects) becomes a source of additional energy output for these values. This contravenes the law of energy conservation. Therefore, we should believe that the density of grain boundaries stops decreasing when zero value is achieved.

If the strain value quickly increases, the density of grain boundaries also increases. This is shown by curve~2, in figure~\ref{fig5}. The grain sizes, in the limiting structures, decrease. This curve has three intersection points with the abscissa axis. The first and the third intersections correspond to stable steady-states [the maxima of the potential $V\left( h_g\right)$ (\ref{eq12})]. However, it should be noted that the first intersection lies in the range of negative values of the density of grain boundaries like in the previous instance. That is why it is supposed that the process of grain scattering slows down. The value of $h_{g0}$ should be assumed to be zero because the values of $h_g<0$ have no physical meaning. Thus, two limiting structures having large and small grain sizes are formed. The second intersection corresponds to the minimum of the potential and, consequently, to the unstable value of $h_g$. The system does not return after the deviation from this state. It is noteworthy that the kinetics  approaching the steady-state is exponential because the $\dot{h}_g$ vs $h_g$ dependence near the intersection point is close to linear.

If we continue to increase $(\varepsilon_{ij}^\mathrm{e} )^2$, three intersection points of the dependence with the abscissa axis are observed. This is shown by curve~3, in figure~\ref{fig5}. However, the main difference is that the first stationary point realizes in the positive range of the density of grain boundaries $h_g$. The first intersection describes the stable steady-state. The limiting structure with large grain sizes is realized since it is close to the value of $h_g=0$. The potential has one minimum and two maxima in the positive range of $h_g$. In the negative range, it monotonously increases for $h_g\rightarrow -\infty$. There is only one steady state at the further $(\varepsilon_{ij}^\mathrm{e})^2$ increase (the curve~4). It corresponds to the realization of a limiting structure having small grain sizes.

It should be noted that in the real experiment  the achievement of a stationary mode is impossible due to the geometric feature of the SPD methods. Such classical methods as equal channel angular pressing, twist extrusion, equal channel multiangular hydroextrusion and so on are characterized by conservation of the sample section before and after the SPD process. However, it is not possible to reach a steady mode because it is impossible to provide elastic strain constant in time  (in our instance $\varepsilon_{ij}^\mathrm{e}$) in these methods. The sample with some constant velocity enters an active zone of the experimental setup and leaves it. The working shear stresses $\sigma_{ij} (i\neq j)$ are changed from zero to some maximum value at the entrance to an active zone and from the maximum value to zero at the exit from it. Nevertheless, in an active zone, the structure of the material will evolve towards the stationary value, but during the sojourn time in the zone, the system will be capable of passing only a part of the way in this direction. Therefore, repeatedly iterative processing of the material on the same setup or a combination of processing on setups of various types are widely practiced to achieve the stationary limiting average grain sizes \cite{Pogrebnjak_V2001,prd,Pogrebnjak_PMM_2011}.

The idea of intensive homogeneous loading of the sample, in the process of its processing, is also used. In this instance, the stationary mode can be achieved by the strain $\varepsilon_{ij}^\mathrm{e} = \text{const}$, but due to the cross-section change of the sample during such processing only for a limited period of time.

\subsection{Check of adiabatic approximation}\label{subsec:level4_3}

The kinetics analysis from the arbitrary initial nonequilibrium state to one or two (at their existence) stationary states was carried out above. In the presence of two stationary states, the first of them describes the structure with the large grain (smaller or zero values of $h_{g0}$) and the second one represents the structure with the fine grain (large values of $h_{g0}$). The transition between stationary states was not considered in this article. This transition presents the essence of grain grinding during the process of severe plastic deformation, at the same time. Such an approach has been investigated in detail earlier in the works of one of co-authors \cite{Metlov_BulletinSerA2009,Metlov_PhysRevE2014}.

This is of interest to consider as far as the hypothesis of an adiabatic process of the defects generation leads to a deviation from the ``exact'' solution. To this effect, the calculation of the density of dislocations and grain boundaries is carried out in the course of evolution {by exact formulas~(\ref{eq7}),~(\ref{eq8}) at model parameters, presented in \cite{Metlov_PhysRevE2014}. Then, the adiabatic value of the dislocations density is calculated by equation~(\ref{eq9}) through exactly finding the solution for grain boundaries. The maximum shear elastic strain $\varepsilon^\mathrm{e}= \mathrm{max}\{\mathrm{shear}(\varepsilon_{ij}^\mathrm{e})\}$ is specified by Taylor relation \cite{Metlov_PhysRevE2014}
  \begin{equation}\label{b12}
\varepsilon^\mathrm{e}=A\sqrt{h_{D}}\,,
  \end{equation}
where a constant $A=2.75\times 10^{-11}$~$\mathrm{m}^{3/2}$.}

\begin{figure}[!t]
       \centerline{\includegraphics[width=0.54\textwidth]{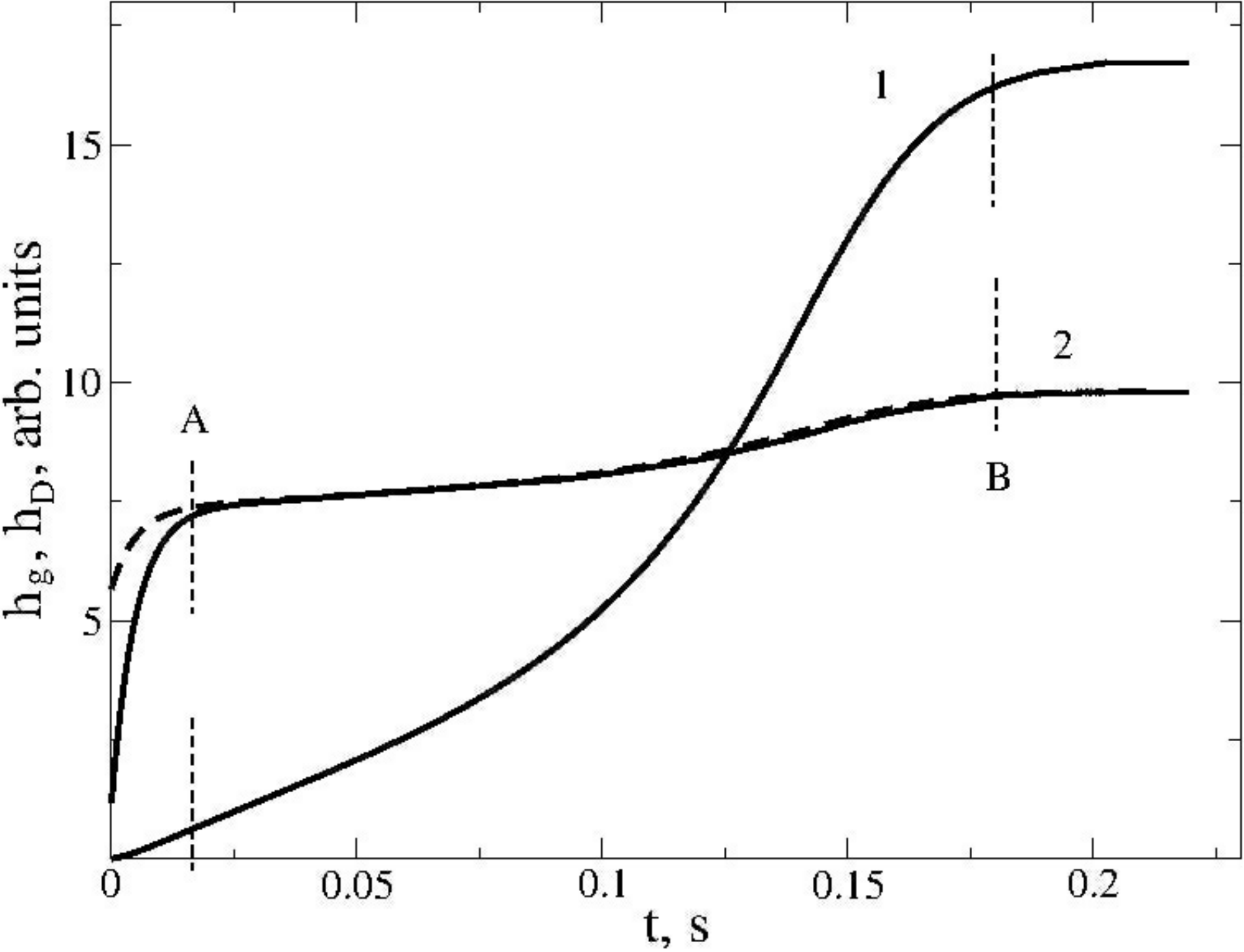}}
       \caption{ The kinetics of the density of grain boundaries $(1)$ and dislocations $(2)$. Solid lines correspond to the ``exact'' solution, dashed line corresponds to the adiabatic approximation. The section to the left-hand side of the point $A$ meets a stage of intensive generation of dislocation, the section $A$--$B$ corresponds to the SPD stage.}
       \label{fig6}
\end{figure}

The difference in the found values defines the degree of deviation from an adiabaticity in this instance (figure~\ref{fig6}). As is seen from the figure, on the whole, the adiabatic approximation for the fixed parameters of the model is performed rather well. The adiabatic solution has the greatest deviation on the initial section of the intensive generation of dislocation. This means that adiabatic approach can be successfully applied, at least, in some instances to the kinetics research of the structural defects in the SPD course.

\section{Conclusions}\label{sec:level5}

The study based on the principles of the Landau theory of phase transitions is presented. This approach enables us to describe the mode of the material fragmentation during the SPD process. The two-defect two-mode model of  nonequilibrium evolution thermodynamics, in the expressions of the internal energy, is assumed to be the basis. The grain boundary and dislocation are chosen in the capacity of the main structural defects.

\looseness=-1This approach allows us to describe the existence of a limiting grain structure (non-zero maximum of the thermodynamic potential) achieved as a result of the SPD process. The coarse-grained state of the material (in a limit single crystal) meets the zero energy maximum and it is examined as a limiting structure, which is equilibrium relatively to the ordinary plasticity in the theory context. It is shown that the transition from a coarse-grained structure to the fine-grained one during the SPD process can take place according to the schemes of the first- and second-order phase transitions. The phase diagram is built, where the values of the first two invariants of the elastic part of the strain tensor $\varepsilon_{ii}^\mathrm{e}$ and $(\varepsilon_{ij}^\mathrm{e} )^2$ define the domains of the realization of various types of the limiting structures. Four domains are allocated, among which two domains with one limiting structure and other two regions with two structures. In these domains, the limiting structures can be formed both for zero value of the density of grain boundaries (large single-crystal grains)  and for non-zero value of the density of grain boundaries (finer grains). The maxima of the potential $V\left( h_g \right)$ correspond to the formation of limiting structures with the different grain sizes.

The kinetics of the steady-state values establishment of the density of defects is investigated. The relaxation dependencies $h_g(t)$ are built for the analysis of the Landau-Khalatnikov equation. The research is carried out for all domains of the phase diagram within the adiabatic approach (at which the dislocations density change follows the evolution of the density of grain boundaries). It is found that the type of the formed limiting structure depends on the initial values of the density of grain boundaries. It is shown that grain sizes, in the limiting structures, decrease with an increase of the elastic strains.

It is possible to predict the existence  probability for the other limiting structures with finer grain sizes in the materials, if we take into consideration higher powers of the expansion of thermodynamic potential. It is necessary to sharply increase the deformation rate (to increase $\varepsilon_{ij}^\mathrm{e}$) in order to experimentally detect  the limiting structures of a higher rank, when the first limiting structure has been attained. This can be achieved either by an increase of the rate of the material fed in the SPD setup, or by a sharp decrease of the cross-section of the blank in the production process with the same feed rate.

\vspace{-2mm}

\section*{Acknowledgements}

The work was executed under support of the Ministry of Education and Science of Ukraine within the framework
of the project ``Nonequilibrium thermodynamics of metals fragmentation and friction of spatially nonhomogeneous
boundary lubricants between surfaces with nanodimensional irregularities'' (No.~0115U000692).


\clearpage

\ukrainianpart

\title{Термодинаміка і кінетика фрагментації твердих тіл\\ при інтенсивній пластичній деформації}
\author{О.В.~Хоменко\refaddr{label1}, Д.С.~Трощенко\refaddr{label1}, Л.С.~Метлов\refaddr{label2,label3}}
\addresses{
\addr{label1} Сумський державний університет, вул. Римського-Корсакова, 2, 40007 Суми, Україна
\addr{label2} Донецький фізико-технічний інститут ім.~О.О.~Галкіна НАН України,
просп. Науки, 46, 03028 Київ, Україна
\addr{label3} Донецький нацiональний унiверситет, вул. 600-рiччя, 21, 21021 Вiнниця, Україна
}

\makeukrtitle

\begin{abstract}
\tolerance=3000%
Розвинено раніше запропонований підхід нерівноважної еволюційної термодинаміки, котрий допомагає в рамках адіабатичного наближення описати процеси дефектоутворення. Базова система рівнянь залежить від початкового розподілу дефектів (дислокацій та меж зерен). Побудована фазова діаграма, що визначає області реалізації різних типів граничних структур. Досліджено взаємодію декількох видів дефектів на формування граничної структури з точки зору внутрішньої енергії. Знайдено умови формування двох граничних структур. В рамках адіабатичного наближення, при якому зміна щільності дислокацій слідує за еволюцією щільності меж зерен, досліджено кінетику встановлення стаціонарних значень щільності дефектів. Встановлено, що зі збільшенням пружних деформацій зменшуються розміри зерен у граничній структурі.
\keywords межа зерна, дислокація, фазовий перехід, гранична структура, внутрішня енергія

\end{abstract}

\end{document}